\documentclass[epsfig,12pt]{article}
\usepackage{epsfig}
\usepackage{amsfonts}
\begin {document}

\title {ON THE SOLUTIONS OF INFINITE SYSTEMS OF LINEAR EQUATIONS}

\author{J.L. Hern\'andez-Pastora\thanks{E.T.S. Ingenier\'\i a Industrial de
B\'ejar. e-mail address: jlhp@usal.es. IUFFyM: Instituto Universitario
de F\'\i sica Fundamental y Matem\'aticas.}  \\
Departamento de Matem\'atica Aplicada and IUFFyM\\
Universidad de Salamanca.  Salamanca,
Spain  \\}

\date{\today}

\maketitle

\begin{abstract}
New theorems about the existence of solution for a system of infinite linear
equations with a Vandermonde type matrix of coefficients
 are proved. Some examples and applications of these results are shown. In
particular, a kind of these systems is solved and  applied in the 
 field of the General Relativity Theory of Gravitation. The solution of the
system is used to construct a relevant physical representation of certain 
 static and axisymmetric solution of the Einstein vacuum equations. In addition,
a newtonian representation of these relativistic solutions 
is recovered. It is shown as well that there exists a relation between this
application and the classical Haussdorff moment problem.
\end{abstract}


\newpage

\section{Introduction}

We study in this work the existence of solution for
 an algebraical system with an  infinite number of
 linear equations and unknowns whose matrix of coefficients is a Vandermonde
matrix. This kind of system of equations
broadly arises, for example, in the field of plasma physics \cite{porras} and
research on this topic becomes
of interest for both a theoretical point of view as well as  experimental
problems in plasma or nuclear physics. In this work we apply our results into
the field of exact solutions in the
General Relativity (GR) Theory of Gravitation where a relevant representation of some
static and axially symmetric solutions of the
Einstein vacuum field equations can be obtained by means of solving that kind of
systems of linear equations.

In the mathematical literature already exists from the nineteenth century the
interest in some cases where an infinite system of linear equations needs to
be solved \cite{riesz}. A method so called {\it the finite section method} is
introduced in the work of Fourier (cited in \cite{riesz}). This method is
an approach to finding a solution of an infinite system of linear equations. A
rigorous and robust analysis of the problem has been developed in \cite{sereny},
as well as previous works have enquired this topic \cite{enqu}. Is not
our aim to improve their results but to use these mathematical techniques  into
a
theoretical physical problem. In \cite{sereny} the authors study how the
infinite section method works for the class of systems described by
an infinite Vandermonde matrix. Here we provide  the problem with alternative
theorems proving the existence of solutions of such kind of
 systems of linear equations
 for a determined set of the independent terms of the system whenever the
coefficients of the Vandermonde matrix fulfill some condition.
These results are
applied to  some toy models as well as to solve an infinite system of linear
equations that appears at
the field of relativistic exact solutions of Einstein equations.

In \cite{dumbel} the  aim of the work consists of looking for some object whose
newtonian gravitational potential reproduces the metric function of some  Weyl
solutions \cite{weyl}. The solution with spherical symmetry of the Einstein
vacuum equations is given by
the Schwarzschild space-time. The relevant metric function  of the line element
of this solution can be interpreted as the
gravitational potential of a bar of length $2M$ ($M$ being the mass of the
solution) with constant linear density $\mu=1/2$.
In \cite{dumbel} we  proved, by carrying on with the spherical-symmetry analogy,
that it is
 possible to construct a well-behaved linear density of a bar, amended with
other characteristics, which allows us to provide a physical interpretation of
some Weyl solutions. In particular the LM solutions, with very interesting
physical properties (see \cite{lmgeod}), is developed.
An alternative way to obtain a  definition of the density given in 
\cite{dumbel}
comes from a discrete outlook of the bar by considering it as the infinite sum
of point-like particles. Hence, we can identify the infinite sum of Curzon-like
potentials \cite{curzon} associated to these masses placed along the bar, with the potential
of the bar obtained by means of  an integral involving the linear density. We
make use of
a limit, where the number of points of the discrete version of the bar goes to
infinity in such a way that  the Riemann  integral sums  lead  to the Leibnitz
integral. Hence
 a continuous definition of the bar is recovered. The matching between both
discrete
and continuous descriptions of the bar  allows us to supply the two families of
parameters appearing at the discrete version of the bar  with a physical
meaning, as well as to
construct a linear density  for the continuous bar.

 Summarizing, we could say that determined Weyl solutions may be  written as a
linear combination
of infinite Curzon solutions and they can be interpreted by means of an
artificial object whose
 gravitational potential provides the metric function of the
solution. Now, the unsolved question is to know whether any solution of the Weyl
family can be written as an infinite sum of Curzon-like potentials bounded by
a finite length   along the symmetry axis. The solution of this
problem is equivalent to solve an algebraical system with an  infinite number of
 linear equations whose matrix of coefficients is a Vandermonde type and its
matrix of independent terms consists of  the set of Weyl coefficients. The
theroems proved in this work try to sed light into this problem.

The work is organized as follows:
Section 2 is devoted to introduce the mathematical problem to solve and some
theorems are proved. Section 3 contains some examples and applications; first,
we
 deal with two different types of successions for the set of independent
coefficients of the lienar system of equations. Second, we apply the theorems
 into the field of GR: when trying to
write an static and axisymmetric solution of the Einstein vacuum equations
 as an infinite sum of Curzon solutions \cite{curzon}, a system of linear equations arises. In
addition we recover a newtonian representation of some relativistic 
solutions by means of the linear density of certain object \cite{dumbel}. We
show that  a discrete version of the problem leads to  a characterization of an 
infinite system of linear equations that can be solved from the continuous
definition of the gravitational potential defined in terms of the density.
The LM \cite{dumbel} and Erez-Rosen \cite{erez} solutions are addressed, and
differences with respect to the spherical case (Schwarzschild solution) are
discussed.

Section 4 is devoted to show that there exists a relationship between the
previous problem and the classical Hausdorff moment problem. A section with
conclusions follows this section and finally an appendix that contains
some
 classical  known results about newtonian gravity, which are used in the text,
is added.

\section{Existence of solution for an infinite system of linear equations}

Let us consider  a system of $n$ linear equations, $A = {\cal V}_n Y$, whose
matrix of coefficients  is given by a Vandermonde matrix,
\begin{equation}
 {\cal V}_n\, =\, {\cal V}_n(x_1,\dots,x_n)\, =\, \left( \begin{array}{cccc}
1&1&\dots &1\\ x_1&x_2&\dots &x_n\\ x_1^2&x_2^2&\dots &x_n^2 \\ \vdots &\vdots &
\ddots &\vdots \\ x_1^{n-1}&x_2^{n-1} &\dots &x_n^{n-1} \end{array}\right)\quad
; \qquad \left({\cal V}_n\right)_{ij}\, =\, x_j^{i-1}
\end{equation}
where  $x_1,\dots,x_n$ is a set of  $n$ real (or complex) numbers, associated to
the Vandermonde matrix, such that $x_i\neq x_j$, $\forall i\neq j$, and $A$
represents the $(n\times 1)$-matrix of independent terms $\left(A\right)_{i}\,
=\, a_i$, and $Y$ represents the $(n\times 1)$-matrix of unknown variables
$\left(Y\right)_{i}\, =\, y_i$.

Since the Vandermonde matrix is invertible the solution of the above mentioned
system of linear equations is unique for any dimension $n$ of the system. The
question is whether we also can obtain solutions of that system  if its
dimension
$n$ goes to infinity. The finite section method \cite{sereny} works as follows:
we consider the first $n$ equations and $n$ unknowns and we solve this finite
 system, where the rest of the terms have been neglected. As $n$ grows larger
these solutions are expected to approximate a solution of the infinite system.
In what follows we present a theorem that prove the convergence of this
procedure if the sets of independent terms and  Vandermonde parameters
 fulfill some particular conditions.

\vspace{3mm}

{\bf Theorem 1: A sufficient condition}

{\it The solution of a system of infinite linear-equations with a Vandermonde
matrix of coefficients exists if the Vandermonde parameters $\{x_i\}$ fulfill
the conditions $x_i\neq x_j$, $\forall i\neq j$, $x_i=1$ for some $i$, and the
set $\{a_i\}$ of  independent terms  of the system is a  bounded, monotonic
succession (and therefore convergent) with $a_i \neq 0 \ \forall i$.}

\vspace{2mm}

{\bf Proof:}

If we write explicitly the unknown variables in terms of
the inverse of the Vandermonde matrix we obtain the following expression:
\begin{equation}
y_k= \sum_{i=1}^{\infty} b_i^{(k)} a_{i-1} \, , \qquad k=1..\infty
\label{serieyk}
\end{equation}
where $b_i^{(k)}\equiv \left({\cal V}_{\infty}\right)^{-1}_{ki}$ represents the
$(ki)$-coefficient  of the inverse Vandermonde matrix of infinite dimension.

The expression (\ref{serieyk}) for the unknown variables is a series whose
convergence is the point we want to prove for any value of $k$. We limit ourself
to consider the independent terms  of the system (the matrix $A$) in such a way
that the set
$\{a_i\}$ be a  convergent succession with $a_i \neq 0 \ \forall i$ . For this
case we make use of the
Dirichlet-Abel theorem that holds the convergence of
the series (\ref{serieyk}) if the series
$\displaystyle{\sum_{i=1}^{\infty}b_i^{(k)}}$ converges as well. The
coefficients $b_i^{(k)}$ can be written as follows
$$
b_i^{(k)}=\frac{v_i^{(k)}}{f_k} \, ,
$$
where  $v_i^{(k)}$ are the coefficients of the following $n-1$ degree
polynomial:
$$
F_j(z)\, =\, \prod_{i=1,i\neq j}^n (z-x_i)=\sum_{i=1}^{n} v_i^{(j)} z^{i-1} \, ,
\, 
$$
and
$$
f_j(x_1,\dots,x_n) \equiv f_j= F_j(x_j)\, .
$$

We can prove that every series $\displaystyle{\sum_{i=1}^{\infty}b_i^{(k)}}$ is
convergent if a suitable choice of the Vandermonde parameters is considered; in
fact, it is easy to see that the notation used leads to the following sum
$S^{(k)}$ of those series:
\begin{equation}
 S^{(k)}\equiv \sum_{i=1}^{\infty}b_i^{(k)}=\frac{F_k(z=1)}{f_k} \, ,
 \label{suma}
 \end{equation}
 i.e., the sum of those series is just the quotient of the values of the
polynomial $F_k(z)$ at the points
$z=1$ and $z=x_k$ respectively. If we consider  the Vandermonde  parameters
  such that one of them is equal to $1$, i.e, $x_i=1$, for $i=i_0$, then we
can  get a convergent sum because the partial sum of order $n$ of the series
(\ref{suma}) are
 \begin{equation}
 S_n^{(k)}\equiv \sum_{i=1}^n b_i^{(k)}=\prod_{i=1,i\neq
k}^n\left(\frac{1-x_i}{x_k-x_i}\right)=\left\{
\begin{array}{ccc} 1&,&k=i_0 : \, x_k=1\\ 0&,&k\neq i_0 : \, x_k \neq 1
\end{array}  \right. \ , 
\label{deltaapp}
 \end{equation}
 and consequently, the series (\ref{serieyk}) are convergent for any $k$ since
the
sums are bounded
$S^{(k)}=\displaystyle{\lim_{n\rightarrow \infty}S_n^{(k)} \neq \pm \infty}$.

{\hfill $\square$}

In this theorem we have forced the succession $\{a_i\}$ to possess no-zero
terms. Nevertheless, it is clear that this condition can be reduced
 to impose that the succession $\{a_i\}$ contains a finite number of zero terms
since the convergence of the series (\ref{suma}) is unchanged in that case.
On the contrary, the special case with a finite number of terms $a_i$ different
to zero needs an alternative treatment.
In particular the case with $\{a_i\}=\{\delta_{j1}\}^{\infty}_{j=1}$ is solved
in \cite{sereny}.

\vspace{2mm}

As a consequence of the previous theorem the solution of the infinite system of
linear equations is given by the following expression:
\begin{equation}
y_k={\displaystyle \frac{\sum_{i=1}^{\infty}
v_i^{(k)}a_{i-1}}{\sum_{i=1}^{\infty}
v_i^{(k)}x_k^{i-1}}} \ .
\label{ydek}
\end{equation}

In the following theorem we introduce an algebraical equation equivalent to the
system of Vandermonde linear equations. The solution
 of this new equation for some toy models, both for the cases with  a  finite or
infinite number of unknowns, is obtained in the following section.

\vspace{4mm}

{\bf Theorem 2: A necessary and sufficient condition}

{\it The solution of a system of $n$ ($\forall n$)linear equations, $A
= {\cal V}_n Y$, whose matrix of coefficients  is given by a Vandermonde matrix,
exists {\bf iff} the following homogeneous differences equation is fulfilled:}
\begin{equation}
\sum_{i=1}^{n+1} v_i a_{i-1}=0 \ ,
\label{lema2}
\end{equation}
{\it where  $v_i$ are the coefficients of the following $n$ degree polynomial:
\begin{equation}
F(z)= \prod_{i=1}^n (z-x_i)=\sum_{i=1}^{n+1} v_i z^{i-1} \ .
\label{lema2w}
\end{equation}
}

\vspace{2mm}

{\bf Proof:}

\vspace{2mm}

A) The necessary condition: Let us suppose that given a succession
$\{a_i\}$ it can be written in the form $A = {\cal V}_n Y$ for some sets of
parameters $\{y_i\}$, and $\{x_i\}$, or equivalently that a succession $\{y_i\}$
verifies the condition (\ref{serieyk}), i.e. a solution of the system of
infinite
linear-equations exists. Therefore, we can  multiply each file $i$ of the system
by $v_i$ and consider the sum of all the files:
\begin{equation}
v_1a_0+v_2a_1+\dots+v_na_{n-1}=y_1\left(\sum_{j=1}^nv_jx_1^{j-1}
\right)+\dots+
y_n\left(\sum_{j=1}^nv_jx_n^{j-1}\right) \ .
\end{equation}
By taking into account that
$\displaystyle{\left(\sum_{j=1}^nv_jx_{a}^{j-1}\right)=-x_a^n}$ since $F(x_a)=0$
for all $a=1..n$, then we can write the following expression
\begin{equation}
\sum_{i=1}^nv_ia_{i-1}=-\sum_{i=1}^nx_i^{n}y_i \ ,
\label{casi}
\end{equation}
and we conclude this part of the proof by means of the starting condition
$\displaystyle{a_k=\sum_{j=1}^nx_j^{k}y_j}$ and the following fact: $v_{n+1}=1$.

\vspace{2mm}

B) The sufficient condition: Let us consider the following differences
equation for the succession $\{a_k\}$:
\begin{equation}
a_{k+n}v_{n+1}+\dots+v_3a_{k+2}+v_2a_{k+1}+v_1a_k=0 \ ,
\label{diffeq}
\end{equation}
or equivalently, we can afford solving the following equation: $p(\lambda)a_k=0$
where $p(\lambda)\equiv \lambda^nv_{n+1}+\lambda^{n-1}v_n+\dots+\lambda v_2+v_1$
is the characteristic polynomial of the differences equation (\ref{diffeq}). The
general solution of this equation is the following:
\begin{equation}
a_k=\sum_{i=1}^n
p_i^k\left(\beta_{i_1}+k\beta_{i_2}+k^2\beta_{i_3}+\dots+k^{m_i}\beta_{i_{m_i-1}
}\right) \ ,
\end{equation}
where $p_i$ are the roots of the characteristic polynomial $p(\lambda)$ with
multiplicity $m_i$ for $i=1..n$ and $\beta_j$ are arbitrary constants. Hence, we
conclude this part of the proof since the roots $p_i$ are the Vandermonde
coefficients $x_i$, and we take
 $\beta_{i_2}=\beta_{i_3}=\dots=\beta_{i_{m_i-1}}=0$, as well as 
$\beta_{i_1}=y_i$ fulfilling the initial
condition $\displaystyle{a_0=\sum_{i=1}^{\infty} y_i}$.

{\hfill $\square$}

\vspace{4mm}

{\bf Theorem 3: Corollary}

{\it With a set of Vandermonde parameters given by  Theorem 1, ($\{x_i\}$
fulfilling the conditions $x_i\neq x_j$, $\forall i\neq j$, $x_i=1$ for some
$i=i_0$) any convergent succession $\{a_n\}$ of positive terms verifies the
following  homogeneous differences equation ($n\rightarrow \infty$)}:
\begin{equation}
\sum_{i=1}^{n+1} v_i a_{i-1} =0\ ,
\label{lemma3.1}
\end{equation}
{\it where  $v_i$ are the coefficients of the following $n$  degree polynomial:
\begin{equation}
F(z)= \prod_{i=1}^n (z-x_i)=\sum_{i=1}^{n+1} v_i z^{i-1} \ .
\end{equation}
}

\vspace{2mm}

{\bf Proof:} It is a consequence of Theorem 1 and Theorem 2.

\section{Some examples and applications}

\subsection{Toy models}

{\bf A) A constant succession}

This kind of succession $\{a_k=a\}$ verifies the following differences equation
\begin{equation}
 a_{k+1}-a_k=0 \ ,
\label{deqA}
\end{equation}
and therefore (\ref{lema2}) implies that $v_1=-1$, $v_2=1$, $v_i=0, \forall
i\neq 1,2$. One solution of the equation (\ref{deqA}) is obtained from the
characteristic polynomial:
\begin{equation}
 (\lambda-1)a_k=0 \Leftrightarrow a_k=1^k c \ ,
\end{equation}
where $c$ is an arbitrary constant. Hence, for the case of a constant
succession, one particular selection of the sets of coefficients $\{x_i\}$ and
$\{c_i\}$ that verify the
condition (\ref{algcond}) are the following
\begin{equation}
 x_{i_0}=1, \quad  c_{i_0}=c=a \quad , \quad x_i=c_i=0 \ \forall i\neq i_{i_0} \ .
\label{conssuc}
\end{equation}
This selection is a degenerate case of the Theorem 1 since $x_i=x_j$, $\forall j
\neq i_0$, and therefore the matrix of the system is not a Vandermonde type
matrix. In fact, the determinant of the matrix of the system is non zero only
for $n=1$.

We give now an alternative argument to get the previous result (\ref{conssuc}):
By virtue of Theorem 2 the condition (\ref{lema2}) is equivalent to the
existence
of the solution
(\ref{ydek}). For the case of a constant succession we find from (\ref{lema2})
that ${\displaystyle a \sum_{i=1}^{\infty}v_i=0}$ which is always true for any
set
$\{x_i\}$ if some point $x_{i_0}$ is equal to $1$, since ${\displaystyle
F(z=1)=\sum_{i=1}^{n}v_i(z=1)^{i-1}=0}$ (see the equation (\ref{lema2w})). In
fact, we know
that (see the equation (\ref{deltaapp}) in Theorem 1)
\begin{equation}
\frac{1}{f_k}\sum_{i=1}^nv_i^{(k)}=\delta_{k,i_{i_0}} \ .
\end{equation}
Let us take for example $\{x_k=k\}$
($i_0=1$, $x_{i_0}=1$) and we find\footnote{The following relations have been
used
\begin{equation}
 f_k=(n-k)!(-1)^{n-k}(k-1)! \ , \ \sum_{i=1}^n v_i^{(k)}=(n-1)!(-1)^{n-1}
\end{equation}
} the following coefficients (\ref{ydek}) $\{c_i\}$:
\begin{equation}
 {\displaystyle c_k=\lim_{n\rightarrow \infty}
\frac{a(-1)^{k-1}(n-k+1)!}{(k-1)!n!}=\delta_{k1}} \ .
\end{equation}

{\bf B) A geometric succession}

This kind of succession $\{a_k=c\eta^k\}$ verifies the following differences
equation
\begin{equation}
 a_{k+1}-a_k \eta=0 \ ,
\label{deqB}
\end{equation}
and therefore (\ref{lema2}) implies that $v_1=-\eta$, $v_2=1$, $v_i=0, \forall
i\neq 1,2$. One solution of the equation (\ref{deqB}) is obtained from the
characteristic polynomial:
\begin{equation}
 (\lambda-\eta)a_k=0 \Leftrightarrow a_k=\eta^k c \ ,
\end{equation}
where $c$ is an arbitrary constant, and hence, for the case of a geometric
succession suitable sets of coefficients $\{x_i\}$ and $\{c_i\}$ that verify the
condition (\ref{algcond}) are the following
\begin{equation}
 x_{i_0}=\eta, \quad c_{i_0}=c=a_0 \quad , \quad x_i=c_i=0 \ \forall i\neq
i_{i_0} \ .
\label{geomsuc}
\end{equation}

 If we alternatively use the equation (\ref{ydek}) to obtain the  set $\{c_i\}$
we find the following conclusion if we take $x_{i_0}=\eta$:
\begin{equation}
 c_k=\lim_{n\rightarrow \infty}
\frac{\sum_{i=1}^nv_i^{(k)}\eta^{i-1}c}{f_k}=c\delta_{ki_{i_0}} \ .
\end{equation}
It is clear that $\eta=1$ ends up with the previous case.

\subsection{Representations of relativistic solutions}

\subsubsection{Static and axisymmetric vacuum solutions.}

As is known, the line element of  a static and axisymmetric   vacuum space-time
is represented in Weyl form as follows
\begin{equation}
ds^2 = -e^{2\Psi} dt^2 +e^{-2\Psi}\left[
e^{2\gamma}\left(d\rho^2+dz^2\right)+\rho^2 d\varphi^2\right] \ ,
\end{equation}
where $\Psi$ and $\gamma$ are functions of the cylindrical coordinates $\rho$
and
$z$ alone. The metric function $\Psi$ is a solution of the Laplace's equation
($\triangle \Psi=0$), and the other metric function $\gamma$ satisfies a system
of differential equations whose integrability condition is just the equation for
the function $\Psi$. The Weyl family of solutions with  a good asymptotical
behaviour is given in spherical coordinates $\{r, \theta\}$ as the series ($r
\equiv \sqrt{\rho^2+z^2}$, $\cos\theta= z/r$)
\begin{equation}
  \Psi=\sum_{k=0}^\infty\frac{a_k}{r^{k+1}}P_k(\cos\theta) \ .
\label{psi}
\end{equation}

The Weyl
series (\ref{psi}) could be rewritten as the following linear superposition
of Curzon solutions:
\begin{equation}
{\displaystyle{
\Psi=\sum_{k=0}^{\infty}\frac{a_k}{r^{k+1}}P_k(\cos\theta)=\sum_{k=1}^{\infty}}
\frac{c_k}{\sqrt{\rho^2+(z-z_k)^2}}} \ ,
\label{ocho}
\end{equation}
where the  Curzon solution \cite{curzon}
corresponds to  a point-like particle with mass $c_k$ located at the point $z_k$
on the $Z$ axis. The equation (\ref{ocho}) for the metric function $\Psi$
requieres the following relation between Weyl coefficients and the parameters
$c_i$, $z_i$:
\begin{equation}
 a_k=\sum_{i=1}^{\infty} z_i^k c_i  \ , \ \forall k \ .
\label{algcond}
\end{equation}

The expression (\ref{algcond}) for $a_k$ can be seen as an algebraical
condition written in
matrix form  as follows:
\begin{equation}
\left(
\begin{array}{c}
a_0\\
a_1\\
a_2\\
\vdots
\end{array}
\right)
=
\left(
\begin{array}{cccc}
1&1&1&\dots\\
z_1&z_2&z_3&\dots\\
z_1^2&z_2^2&z_3^2&\dots\\
\vdots&\vdots&\vdots&\dots
\end{array}
\right)
\left(
\begin{array}{c}
c_1\\
c_2\\
c_3\\
\vdots
\end{array}
\right)\ .
\label{matrices}
\end{equation}

The question previously formulated is
whether any set of coefficients $\{a_k\}$ can be written in that form
(\ref{algcond}), or equivalently we want to know whether every solution of the
Weyl family
could be written in the form (\ref{ocho}). Hence we are looking for  sets of
values $\{z_k\}$, $\{c_k\}$ corresponding to certain set
of known values $\{a_k\}$. Equation (\ref{matrices}) represents an infinite
system of
linear equations that is invertible at any order if
the matrix of coefficients possesses a Vandermonde determinant which is
different from zero {\it iff} $z_i
\neq z_j$, $\forall i\neq j$. In the previous section we have proved some
theorems  that show the existence  
of a convergent solution of this
system with an infinite number of linear-equations:  a suitable choice of the
Vandermonde parameters
$\{z_k\}$, if  the coefficients $\{a_k\}$  generate a
convergent succession with a finite number of null terms, is a sufficient
condition.

The toy models that we have previously studied correspond to a particular set
of Weyl coefficients. From those results, we already can state that the solution
of the Weyl family whose coefficients $\{a_k\}$ are of all them equal to a constant, can
be
represented by a single point-like particle of mass $c$ displaced from the
origin of coordinates. Let us note that this
is a known result
because a single point-like mass at the position $z=1$ is described  with the
following expression:
\begin{equation}
 \Psi=\frac{-c}{\sqrt{\rho^2+(z-1)^2}} \ ,
\end{equation}
and this  metric function can be written in the Weyl
form, by means of the decomposition of that function in terms of Legendre polynomials series, as follows ($\omega \equiv \cos \theta$):
\begin{equation}
\Psi= \sum_{k=0}^{\infty}\frac{a_0}{r^{k+1}}P_k(\omega) \quad , \  a_0=-c \ .
\end{equation}
Obviously, the trivial case of a single point-like particle of mass $c$ located at the origin of coordinates (the original Curzon solution), i.e., 
$\Psi=\displaystyle{\frac{-c}{\sqrt{\rho^2+z^2}}}=\displaystyle{\frac{a_0}{r}P_0(\omega)}$ corresponds to the set of coefficients $\{c_1\equiv c=-a_0, c_i=0,  \forall i\neq1\}$, $\{z_1=0\}$.

In addition, from the  example {\bf B)} in section $3.1$ we can say that the
solution of the Weyl family whose  coefficients are given by a geometric
succession $\{a_k=c \eta^k\}$ is
represented by a single point-like particle of mass $c$ located at the point
$z=\eta$.

The requirement imposed on the sets of coefficients $\{a_k\}$, $\{c_k\}$,
$\{z_k\}$ by equation (\ref{algcond}) supply us with a gauge of freedom
 for the selection of the coefficients  since we are dealing with  two sets of
arbitrary
 parameters in the second series of the equation (\ref{ocho}) instead of the unique set of Weyl
coefficients. In the following subsection we address the resolution of the
equation (\ref{algcond}) by means of an infinite set of coefficients $\{z_i\}$,
for a class of static and axisymmetric relativistic solutions.

\subsubsection{A Newtonian representation of the Weyl solutions.}

In \cite{dumbel} a description of some relativistic solutions by means of a
singular Newtonian source is developed. The argument used is the equivalence
between the newtonian gravitational potential $\Phi$ and the metric function
  $\Psi$ of the relativistic solution.
An object so called  {\it dumbbell} is constructed to describe a class of
static and axisymmetric solutions. This object consists of a bar of length $2M$
with determined linear density ($M$ being the mass of the relativistic solution)
and a ball at each end of the bar. The density of the bar is used to describe
physical properties and relevant characteristics of the relativistic solution.
In the Appendix we address a brief review about Newtonian Gravity contents. Some
formulae that we shall use in what follows are introduced.

The aim of this section is  approximating a bar, endowed with a
continuous line density, with a series of point-like masses. The
corresponding
gravitational potential could be written as a linear combination of Curzon
solutions. In other words, we want to search for a set of coefficients $\{c_i\}$
 in such a way that  a linear superposition of Curzon solutions with respective
masses $c_i$ placed in points $z_i$ along the $Z$-axis provides
  a potential $\Psi$ like (\ref{ocho}), as well as  the relation
(\ref{algcond}) could be  satisfied.

We are constrained by the fact that our discrete
 representation of the bar is requested to recover the linear density of this
object. The set of coefficients $\{z_i\}$ should provide us with  a
 continuous distribution of points along  the $Z$ axis between  both ends of the
bar. The transition from the  integral expression (A4)
 to a Riemann integral sum is made by means of a limit that makes the width
of the sub-intervals of integration  goes to zero.
Consequently the number of points  of the partition on the interval $[-L,L]$
tends to infinity in such a way that we handle with the following homogeneous
distribution of $2n$ points:
\begin{equation}
 z_i=\pm i \frac Ln \ , \ x_i \equiv \frac{z_i}{L} \qquad , \qquad i=1 \dots n \
 .
\label{partition}
\end{equation}
Within this selection of coefficients $\{z_i\}$, the solution of the equations
(\ref{algcond})
provide us with  a set of coefficients $\{c_i\}$ representing the masses located
at each point $z_i$.
The discrete consideration of the bar, as well as the
implementation of the condition (\ref{ocho}) allows us  to identify the
coefficients $\{c_k\}$ and supply them with  a physical significance from the
gravitational
potential of the bar (A4). The integral appearing at this equation can
be given  as a limit (Leibnitz notation)
of the Riemann integral sums:
\begin{equation}
\Phi (\vec x) =  -\int_{-L}^L
\frac{\mu(z^{\prime})}{\sqrt{\rho^2+(z-z^{\prime})^2}}
dz^{\prime}=\lim_{l_{\wp}\rightarrow 0}\sum_{i} K(\wp,f) \triangle z_i \ ,
\end{equation}
where
$f(z^{\prime})\equiv-\frac{\mu(z^{\prime})}{\sqrt{\rho^2+(z-z^{\prime})^2}}$,
$\wp$ denotes a partition of the integration interval, $\triangle z_i$
 is the width between two adjoining points of the partition, $l_{\wp}$ being the
symbol to define the generic width of the partition, and $K(\wp,f)$
represents the maximum or minimum of the function $f$ in each sub-interval.
In fact, the required identification between the Newtonian potential
(A4) and the relativistic metric function $\Psi$ (\ref{ocho}) leads to 
the following interpretation  of the quantities
 ${\displaystyle \frac{c_i}{\triangle z_i}}$: they
must be the maximum or minimum values of the density
$\mu(z)$
in each sub-interval, and hence  the following relation could be fulfilled:
\begin{equation}
\lim_{n\rightarrow \infty}
\sum_{i=1}^n \frac{c_i/\triangle z_i}{\sqrt{\rho^2+(z-z_i)^2}}
\triangle z_i=-\int_{-L}^L
\frac{\mu(z^{\prime})}{\sqrt{\rho^2+(z-z^{\prime})^2}} \ .
\end{equation}
Therefore, an infinite set of  point-like particles withrespective masses $c_i$
defines a collection of 
 Curzon potential whose masses  are infinitesimal quantities characterized by
the 
following way:
\begin{equation}
 \mu(z_i)\equiv -\frac{c_i}{\triangle z_i} \quad  , \quad  \lim_{n\rightarrow \infty, 
l_{\wp}\rightarrow 0} \mu(z_i)=\mu(z)
\delta(z-z_i) \ .
\label{delta}
\end{equation}
Hence the density of the dumbbell bar can be defined as the  continuous
limit
of  a discrete function that is defined at each point $z_i$ of the  bar with
the value
$-c_i/\triangle z_i$.

\vspace{4mm}

{\bf A) The Schwarzschild solution.}

We are going to analyze thoroughly the case of spherical symmetry. For this case
we can prove the following theorem:

\noindent {\bf Theorem}

{\it  With the
partition of the interval $[-L,L]$ given by (\ref{partition}) the Weyl
coefficients
$\{a_k\}$ of the Schwarzschild  solution  fulfill the condition
(\ref{algcond}), when $n$ goes to infinity, if ${\displaystyle
c_i=-\frac{L}{2n}}$, $\forall i$.}

\vspace*{5mm}

\noindent Proof:

Accordingly to (\ref{algcond}) we have to prove the following condition:
\begin{equation}
 a_k^{schw}=-\frac 12 \lim_{n \rightarrow \infty}  \left(\frac
Ln\right)^{k+1}\left[ \sum_{i=1}^{n} i^k+\sum_{i=1}^{n} (-i)^k\right] \ .
\label{I}
\end{equation}

Firstly let us note that the first equation in (\ref{algcond}) ($k=0$) implies
the following constraint:
\begin{equation}
 \sum_{i=1}^{\infty}c_i=a_0^{schw} \Leftrightarrow \lim_{n \rightarrow \infty}
\sum_{i=1}^{2n}c_i=-M \ ,
\end{equation}
which is verified by the formulated coefficients $c_i$ if we take $L=M$.
Secondly,
it is also clear from (\ref{I}) that odd coefficients $a_{2k+1}^{schw}$ are zero
because of the equatorial symmetry ($|z_i|=i L/n$).
And finally, in the general case $k=2j$ we have
\begin{eqnarray}
 a_{2j}^{schw}&=&-\lim_{n \rightarrow \infty}  \left(\frac Ln\right)^{2j+1}
\sum_{i=1}^{n} i^{2j}=\nonumber\\
&=&-\lim_{n \rightarrow \infty}  \left(\frac Ln\right)^{2j+1}
\left[\frac{n^{2j+1}}{2j+1}+ \frac 12 n^{2j}+O(n)^{2j-1}\right] \ ,
\label{II}
\end{eqnarray}
where $O(n)^{2j-1}$ denotes all terms\footnote{These factors are given by the
Bernoulli numbers (see \cite{abramo} for details).} with powers of $n$ less than
$2j$.
Hence equation (\ref{II}) shows that Weyl coefficients for the Schwarzschild
solution are ${\displaystyle a_{2j}^{schw}=-\frac{L^{2j+1}}{2j+1}}$ for all
$j\geq 1$,
and so we recover from the continuous limit (\ref{delta}) the known result for
the density of the bar in this case:
\begin{equation}
 -\frac{L}{2n}\equiv c_i \ : \ \mu(z_i)=-\lim_{n \rightarrow \infty} \frac{c_i
}{L/n}
\Leftrightarrow \mu(z_i)=\frac 12 \quad , \quad  \forall i \ ,
\end{equation}
where we have used $\triangle z_i\equiv dz=z_{i+1}-z_i=\frac Ln$.

$\hfill{\square}$

\vspace{2mm}

{\bf B) The Erez-Rosen {\it vs}  the M-Q$^{(1)}$ solution.}

These  are both two-parameters non spherical symmetric solutions of Weyl family.
One of the parameter $M$
represents the mass and the other $q$ or $q_2$ denoting the
dimensionless quadrupole moment for M-Q$^{(1)}$ or ER solutions
respectively, and related by $q_2=\frac {15}{2} q$ \cite{tesis}. In
\cite{luisgyr} a comparison between
both solutions were done and different conclusions  were obtained regarding the
behaviour  of gyroscope precessing in circular orbits into these
 gravitational fields. The Weyl coefficients of the ER solution are known
\cite{tesis}:
\begin{equation}
 a_{2k}^{ER}=-\frac{M^{2k+1}}{2k+1}\left(1+q_2\frac{2k}{2k+3}\right) \ ,
a_{2k+1}^{ER}=0 \ ,
 \label{anER}
\end{equation}
and a representation of this solution by means of a bar with linear density
\begin{equation}
 \mu^{ER}(X)=\frac 12 \left(1-\frac{q_2}{2}\right)+\frac 34 q_2 X^2 \ , \  X
\equiv
\frac{z}{M} \ \in \left[-1,1\right] \ ,
 \label{densiER}
\end{equation}
is carried out in \cite{dumbel}.
We can solve the equations (\ref{algcond}) for the set of coefficients $\{c_i\}$
assuming that $|z_i|=iL/n$ in analogy with the spherical case (Schwarzschild's solution).
It is easy to
prove that
\begin{equation}
 c_i=-\left( 1-\frac{q_2}{2}\right) \frac{L}{2n}-\frac
34\frac{q_2}{M^2}\left(\frac Ln\right)^3 i^2 \ ,
\end{equation}
since these quantities verifies ($L=M$)
\begin{eqnarray}
 a_{2k}&=&\lim_{n\rightarrow\infty}\sum_{k=1}^{n}z_i^{2k}c_i=\nonumber \\
&-&\frac 12\lim_{n\rightarrow\infty}\left[\left(1-\frac{q_2}{2}\right)
\left(\frac Ln\right)^{2k+1}\sum_{i=1}^{n}(\pm i)^{2k}-\frac
32\frac{q_2}{M^2}\left(\frac Ln\right)^{2k+3}
\sum_{i=1}^{n}(\pm i)^{2k+2}\right]=\nonumber\\
&-&\left(1-\frac{q_2}{2}\right)\frac{L^{2k+1}}{2k+1}-\frac 32 \frac{q_2}{M^2}
\frac{L^{2k+3}}{2k+3}=a_{2k}^{ER} \ .
\label{aesER}
\end{eqnarray}

In addition we see that these selection of coefficients $\{z_i\}$, $\{c_i\}$
allows us to recover the density (\ref{densiER}) by satisfying the 
condition (\ref{delta}):
\begin{equation}
\frac{c_i}{\triangle z_i}=-\frac 12 \left( 1-\frac{q_2}{2}\right) 
-\frac34 q_2\left(\frac{iL}{nM}\right)^2=-\mu(X_i) \ .
\end{equation}

\vspace{2mm}

In contrast with the ER solution, the M-Q$^{(1)}$ solution is represented by a
dumbbell consisting of a bar of length $2L$ and linear density (see
\cite{dumbel})
\begin{equation}
\mu^{MQ^{(1)}}(X)=\frac 12 \left(1-\frac{15}{8}q\right)+\frac{15}{16}q X^2 \ ,
\end{equation}
and a point-like particle  at each  end of the dumbbell with
respective mass ${\displaystyle \nu=\frac 58 qM}$.

We can solve the system of equations (\ref{algcond}) for the unknown set of
coefficients $\{c_i\}$ in a similar way used for the Schwarzschild solution
(\ref{I}-\ref{II}). We suppose again an homogeneous distribution of points
$\{z_i\}$ recovering the bar from one end to another (\ref{partition}). The
M-Q$^{(1)}$ solution is a subclass of the 
LM solution \cite{dumbel}, and for all these solutions we can make use of the
following decomposition of the coefficients\footnote{In particular, the Weyl
coefficients of the  M-Q$^{(1)}$ solution are ${\displaystyle
a_{2k}=-\frac{M^{2k+1}}{2k+1}\left(1+q\frac{5k(k+2)}{2k+3}\right)}$ and they can
be decomposed as follows: ${\displaystyle
a_{2k}=-\frac{M^{2k+1}}{2k+1}\left(1-\frac{15q}{8}\right)-\frac{M^{2k+3}}{2k+3}
\left(\frac{15q}{8M^2}\right)-M^{2k+1}\frac 54 q}$.}
$\{a_k\}$:
\begin{equation}
a_{2k}^{LM}=-\sum_{j=0}^g\frac{M^{2k+2j+1}}{2k+2j+1} \frac{H_j}{M^{2j}} -
M^{2k+1} H \ ,
\label{anLM}
\end{equation}
where $H$ and $H_j$ are well defined coefficients (in particular $H_j/2$ are the
coefficients of the density $\mu(X)$ \cite{dumbel}).
Therefore, it is easy to prove that the solution of (\ref{algcond}) for this
case must be
obtained by splitting the expression (\ref{anLM}) in two parts since it
provides two kind of successions for $\{a_n\}$.
On the one hand, with respect to the second term in (\ref{anLM}) we must
consider one point-like mass located at each end of the bar.
The general term of the succession associated to
the second part of (\ref{anLM}) $\{-M^{2k+1} H\}=\{\eta^{2k} b\}$ provides a 
geometric
type succession 
with rate $\eta\equiv M$ and initial term $a_0\equiv b = -M H$, and therefore
the
parameter $-b/2$ acquires the meaning of being the mass of
each point-like particle located at both ends of the bar\footnote{An alternative
argument comes from the direct identification
between the gravitational potential of two particles of mass $m$ situated at
distances $-L$ and $L$ respectively along the $Z$ axis, ${\displaystyle \Psi
=\Phi=
-\frac{m}{\sqrt{\rho^2+(z+L)^2}}-\frac{m}{\sqrt{\rho^2+(z-L)^2}}}$, with the
corresponding metric function by means of performing a power
series expansion as follows: ${\displaystyle \Phi=
-\frac{m}{\sqrt{\rho^2+(z+L)^2}}-\frac{m}{\sqrt{\rho^2+(z-L)^2}}=
-\sum_{n=0}^{\infty}\frac{P_{2n}(\omega)}{r^{2n+1}} \left(2m L^{2n}\right)}$,
and therefore the Weyl coefficients related to this potential of two particles
are $a_{2n}=-2m L^{2n}$. Let us remind that the condition (\ref{algcond}) is
fulfilled and
so, ${\displaystyle a_0=\sum_{i=1}^{\infty}z_i^0 c_i=-2m}$. That is to say, the
sum of the
coefficients
$\{c_i\}$ corresponding to the geometric succession in (\ref{anLM}) provides the
mass of both
balls except for sign.}.

On the other hand, with respect to the first part of the Weyl coefficients
(\ref{anLM}) is easy to prove, in similar way to the applied for
Schwarzschild solution (\ref{II}) that the corresponding coefficients $\{c_i\}$ of the MQ$^{(1)}$ solution
 fulfilling the equations (\ref{algcond}) are
\begin{equation}
c_i=-\frac 12 \sum_{j=0}^1 \left(\frac Ln\right)^{2j+1} i^{2j}
\frac{H_j}{M^{2j}} \ , \  H_0=1-\frac{15}{8}q, H_1=\frac{15}{8}q \ .
\label{cesLM}
\end{equation}

\vspace{2mm}

Before concluding this section it should be pointed the consistency between
formulae derived from Theorems 1 and 2,  and these results obtained above. Let
us consider the 
notation $\{x_i=z_i/L\}$ and hence the coefficients $\{a_i\}$ of the theorems
transforms into $\{a_i/L^i\}$.
The equation (\ref{ydek}) in previous section allows us to calculate the
coefficients
$\{c_i\}$ for a given set of $\{z_i\}$ and $\{a_i\}$. By taking into account
that $\{a_i\}$ represent the newtonian
 multipole moments of the object whose density has been defined, we can put
expression (A5) into (\ref{ydek}) to obtain:
\begin{eqnarray}
 c_k&=&\frac{1}{f_k}\sum_{i=1}^{\infty} v_i^{(k)}
a_{i-1}=- \int_{-1}^1 L
\mu(Lx)\left[\frac{1}{f_k}
\sum_{i=1}^n v_i^{(k)} (Lx)^{i-1}\right] dx  \nonumber \\
&=&-\int_{-L}^L   \mu(z) \left[\prod_{j=1,j\neq
k}^n\frac{z-z_j}{z_j-z_k}\right] dz \ ,
\end{eqnarray}
and consequently, by taking into account the equation (\ref{deltaapp}), since $z
\in \left[-L,L\right]$ and the Vandermonde parameters $\{z_i\}$ generate an
homogeneous distribution of points from $z=-L$ to $z=L$, then  we have
\begin{equation}
 c_k=-\lim_{l_{\wp}\rightarrow 0} \sum_{i=1}^{\infty}
\mu(z_i)^{max}\triangle z_i \ \delta(z_i-z_k) = -\mu(z_k)\triangle z_k
 \ .
\end{equation}

\section{Relation with the Hausdorff Moment Problem}

In \cite{dumbel} newtonian representations for a class of solutions of the Weyl
family  are obtained by means of an artificial object  so called
dumbbell.
We must remind that not every solution of  the Weyl family can be identified with the
potential of a bar, and so the search for the object  that is able to
describe other
solutions may be a matter of consideration for future works.
Furthermore we might
consider extended newtonian objects,
 rather than singular sources, to obtain the  gravitational potential for each
particular case.

We have seen in previous sections that this newtonian representation can be
connected with a description of those 
solutions by means of  an infinite sum of Curzon solutions. In fact, Theorem 1
and Theoreme 2 provide conditions to be fulfilled by the Weyl 
coefficients in order to write the metric function of the solution $\Psi$ like
equation  (\ref{ocho}).

We should point out that it is possible to obtain general results about
 the existence of an even
density with prescribed moments like those of equation (A5):
\begin{equation}
 M_{2k}=\int_{-1}^1z^{2k}\mu(z)
dz=\int_0^1w^k\left[\frac{\mu(\sqrt{w})}{\sqrt{w}}\right] dw \ ,
\label{hausm}
\end{equation}
 since Hausdorff \cite{hauss} proved a set of necessary and sufficient
conditions
for the existance of a positive
function $f$ with prescribed  half-range {\it moments} $b_n$ in the sense of
equation (\ref{hausm}), 
$b_n=\int_0^1w^n f(w)dw$, that involves  the following inequality conditions: 
\begin{equation}
 0 \leq \sum_{j=0}^{k}(-1)^{j+k} {{k}
\choose {j}} b_{n+j} \ , \ \forall n,k \geq 0 \ .
\label{condhaus}
\end{equation}
We
are referring to the classical problem in analysis
called the Hausdorff Moment Problem \cite{varios}. This condition is equivalent
to say that the sucession of moments is
completely monotonic.
We want to show that the classical problem of Haussdorf is a (continuous)
integral version of the discrete problem outlined by the equation
(\ref{algcond}), and the inequality conditions (\ref{condhaus}) established by
Haussdorf can be recovered by the equation (\ref{lema2}) of the Theorem 2.
It is known that the Weyl coefficients are the newtonian multipole moments 
(A5) of a relativistic solution (see \cite{dumbel} and references
therein). 
Hence we can write the following expression:
\begin{equation}
 a_{2k}=\int_{-1}^1 x^{2k} \mu(x) dx =2 \int_0^1x^{2k} \mu(x) dx \quad , \quad  a_{2k+1}=0 \ ,
\end{equation}
if we deal with an even  linear density $\mu(x)$ that is describing the solution, as is
the case
for the LM solutions \cite{dumbel}. Theorem 2 states that 
the existence of solution for the infinite system of equations (\ref{algcond})
is equivalent to solve the equation (\ref{lema2}) (in the limit $n$ going to
infinity). It is easy to see that if equation (\ref{lema2}) is fulfilled then
the
following expression is true as well:
\begin{equation}
 \sum_{j=1}^{n+1}v_{j}a_{j-1+k}=0 \ , \ \forall \ k, n \ .
\label{h1}
\end{equation}
If we calculate the coefficients $v_{j}$  corresponding to the Vandermonde
parameters $z_i=i L/n$, then the following inequalities are obtained:
\begin{equation}
 v_{i} \leq {n \choose i} (-1)^{n+i} \ ,
\end{equation}
and then we can hold the following relation:
\begin{equation}
 0=\sum_{j=0}^nv_{j+1}a_{j+k}\leq \sum_{j=0}^n (-1)^{n+j+1} {n \choose j+1}
a_{j+k} \ , \ \forall n,k\geq 0 \ .
\end{equation}
This expression reproduces the  Hausdorff condition (\ref{condhaus}) for the
existence of solution of the classical moment problem. Therefore we can
conclude 
that Theorem 2  is an alternative statement for the existence requirements of
the classical Haussdorf problem.

\section{Conclusions}

In GR the static and axially symmetric solutions of the Einstein vacuum
equations are described by a single metric function, which is characterized 
by a set of Weyl coefficients $\{a_i\}$. For some solutions this metric
function can be obtained as the gravitational potential of a newtonian object
\cite{dumbel}.
 A representation of the solution arises from the physical characteristics of
the object, in particular its linear density, and physical properties of the
relativistic solution can be described in terms of this density \cite{dumbel},
\cite{lmgeod}. In this work, we have shown that this linear density can be
connected with a discrete description of the object leading to an infinite sum
of Curzon potentials. We deal with an infinite set of point particles located,
in an homogeneouos distribution, along the symmetry axis into a finite length.

The convergence of this problem is equivalent to the existence of a solution for
an infinite system of linear equations with a Vandermonde type of coefficients
matrix. The matrix of independent terms is given by the set of Weyl coefficients
and the unknows variables are the masses of the point particles. We have proved
some theorems that allow us to guarantee the existence of solution for this kind
of infinite systems. In particular, the solution of this problem for the LM and
Erez- Rosen solutions is proved, by a suitable selection of the Vandermonde
parameters which represent the positions of the point particles along the axis.

Finally, we have proved that a relationship exists between the classical
Haussdorf moment problem and the existence of a newtonian representation of
relativistic solutions by means of an object with a linear density. In
particular the condition held by the Theorem 2 for the existence of solution of
an infinite system of linear equations is equivalent to the condition
established by the Haussdorf moment problem.

\section{Acknowledgments}
This  work  was partially supported by the Spanish  Ministerio de Ciencia e
Innovaci\'on under Research Project No. FIS 2012-30926, and the Consejer\'\i a
de
Educaci\'on of the Junta de Castilla y Le\'on under the Research Project Grupo
de Excelencia GR234. I want to thank Dr. Jos\'e Mar\'\i a  Mu\~noz Casta\~neda 
and Dr. Miguel \'Angel Gonz\'alez Le\'on their
 help in the search for some references.  I  gratefully acknowledges discussions
on this topic with my collagues Dr. D. Miguel \'Angel Gonz\'alez Le\'on, Dr. D.
Alberto Alonso
and professor Dr. D. Juan Mateos Guilarte.

\section{Appendix}

The Newtonian gravitational potential of a mass distribution
with density
$\rho(\vec {\hat z})$, given by the following solution of the Poisson equation
$$
\Phi (\vec x) = - \int_V\frac 1R \mu(\vec {\hat z}) d^3\vec {\hat z} \ , {\hfill (A1)}
$$
where we have used units in which the gravitational constant $G=c=1$, the
integral
is extended to the volume
of the source, $\vec {\hat z}$ is the vector that gives the position of a
generic point inside the source,
and
$R$ is the distance between that point and any exterior point
$P$ defined by its position vector $\vec x$. Let us now make an expansion of
this potential in  a power series of the inverse of the distance from the origin
to the point $\vec P$ ($r \equiv{\cal j}\vec x {\cal j}$) by means of a Taylor
expansion of the term
$\displaystyle{\frac{1}{R}}$ around the origin of coordinates, where $R\equiv
\sqrt {(x^i-{\hat z}^i)(x_i-{\hat z}_i)}$. For the case of an axially symmetric
mass distribution, this multipole development leads to a
Newtonian potential with the same form as equation (\ref{psi}) but the Weyl
coefficients $\{a_k\}$ being replaced by $-M_k^{NG}$, which
are parameters that denotes  the massive multipole moment of order $k$ which can
be
defined by means of an integral expression extended
to the volume of the source,
$$
M_k^{NG} = 2 \pi \int\int {\hat z}^{k+2}\mu({\hat
r},\hat\theta)P_k(\cos\hat\theta)
\sin\hat\theta d\hat\theta d{\hat r} \ , {\hfill (A2)}
$$
${\hat r}\equiv |\vec{\hat z}|$ representing the radius of the integration point
and
$\hat\theta$ the corresponding polar angle.

There is a well-established framework in Newtonian Gravity (NG) for
handling distributional line-sources like a bar of
length $2L$ centered and located along
the $Z$ axis. Therefore we can consider
 an object described by a line singularity on the $Z$ axis with the following
linear density:
$$
 \mu(\vec {\hat z})= \frac{1}{2\pi}\frac{\delta(\hat \rho)}{\hat \rho}\mu({\hat
z}) \ , {\hfill (A3)}
$$
for some non-negative function $\mu(\hat z)$, $\delta(\hat \rho)$ being the
Dirac's
function  $\delta(\hat \rho-\hat \rho_0)$ at $\hat \rho_0=0$ and where  $\{\vec
{\hat z}\}
\equiv \{{\hat \rho},{\hat z}\}$ are cylindrical coordinates.
Consequently, from equation (A1) the gravitational potential of
such a  mass distribution is the
following:
$$
\Phi (\vec x) = - \int_V\frac 1R \mu(\vec {\hat z}) d^3\vec {\hat z} =
-\int_{-L}^L
\frac{\mu({\hat z})}{\sqrt{\rho^2+(z-{\hat z})^2}} d{\hat z}\ , (A4)
$$
where the position vector $\vec x$ is given by coordinates $(\rho,z)$, and $\vec
{\hat z}$ is located along the $Z$ axis.

According to the equation (A2)  the Newtonian multipole
moments of this object (if the function $\mu(\hat z)$ is even in $\hat z$) are
as follows:
$$
 M_{2k}^{NG}=\int_{-L}^L  {\hat z}^{2k} \mu({\hat z}) d{\hat
z}=L^{2k+1}\int_{-1}^1 X^{2k} \mu(LX)dX \ . {\hfill (A5)}
$$

\end{document}